\documentclass[prl,twocolumn,showpacs,preprintnumbers,
  amsmath,amssymb,nofootinbib]{revtex4}
\usepackage{graphicx}
\usepackage{dcolumn}
\usepackage{bm}
\begin{document}
\newcommand{\be}{\begin{equation}}
\newcommand{\ee}{\end{equation}}
\newcommand{\lb}[1]{\label{#1}}
\newcommand{\ca}{{\mathcal A}}
\newcommand{\cu}{{\mathcal U}}
\newcommand{\cK}{{\mathcal K}}
\newcommand{\cT}{{\mathcal T}}
\newcommand{\ptl}{\partial}
\newcommand\lgth{[\,\text{\rm length}\,]}
\newcommand{\Sig}{\Sigma}
\newcommand{\enl}{\\\hfill\rule{0pt}{0pt}}
\newcommand{\sfrac}[2]{\frac{#1}{#2}}
\newcommand{\ct}[1]{\cite{#1}}
\newcommand{\textfrac}[2]{{\textstyle\frac{#1}{#2}}}
\newcommand{\mnote}[1]{\footnote{ MNOTE #1}}

\title{Homoclinic chaos and energy condition violation}

\author{J.\ Mark Heinzle${}^{1}$}
\email{Mark.Heinzle@aei.mpg.de}

\author{Niklas R\"ohr${}^{2}$}
\email{niklas.rohr@kau.se}

\author{Claes Uggla${}^{2}$}
\email{Claes.Uggla@kau.se}

\affiliation{${}^{1}\!$
Institute for Theoretical Physics, University of Vienna, A-1090 Vienna, Austria}

\affiliation{${}^{2}\!$Department of Physics, University of
    Karlstad, S-651 88 Karlstad, Sweden}

\date{July 31, 2006}
\begin{abstract}

In this letter we discuss the connection between so-called
homoclinic chaos and the violation of energy conditions in locally
rotationally symmetric Bianchi type IX models, where the matter is
assumed to be non-tilted dust and a positive cosmological constant.
We show that homoclinic chaos in these models is an artifact of
unphysical assumptions: it requires that there exist solutions with
positive matter energy density $\rho>0$ that evolve through the
singularity and beyond as solutions with negative matter energy
density $\rho<0$. 
Homoclinic chaos is absent when it is assumed that the dust
particles always retain their positive mass.
In addition, we discuss more
general models: for solutions that
are not locally rotionally symmetric we demonstrate that
the construction of extensions through the singularity, which is required for homoclinic chaos,
is not possible in general.

\end{abstract}
\maketitle

The dynamics of locally rotationally symmetric (LRS) and diagonal
Bianchi type IX models with a perfect fluid, usually chosen to be
dust, and a positive cosmological constant has been discussed
recently in a series of
articles~\cite{olietal97,baretal01,olietal02,soastu05}. It was
claimed that Einstein's static universe is a chaotic scatterer so
that ``homoclinic chaos'' is generated in the models. In another
article~\cite{heietal05}, LRS type IX models with general equations
of state (comprising dust and a positive cosmological constant as a
special case) were re-investigated by employing a scale-invariant
formulation. This permitted the use of rigorous dynamical systems
methods which led to results that excluded the possibility of
homoclinic chaos; the dynamics of the models was found to be
predictable. In spite of these rigorous results,
claims about homoclinic chaos persist in the literature.
The purpose of this letter is to resolve these contradictions.

In the first part of this letter
we show that homoclinic chaos in LRS Bianchi type IX models is
an artifact of unphysical assumptions:
for the models to exhibit
homoclinic chaos solutions have to be extended through the
singularity where it must be assumed that the mass of the dust
particles changes sign. The assumption of standard energy conditions
(such as $\rho>0$ for dust) excludes homoclinic chaos and leads to
the results of~\cite{heietal05}.
In the second part, in addition to the issues of energy
condition violation, 
we point
out that the feasibility of the (unphysical)
construction leading to homoclinic chaos
entirely relies on the assumption of LRS symmetry, i.e., extensions of
solutions through the singularity are not possible in more general
models
--- the special structure of LRS type IX models is misleading.

For LRS Bianchi type IX models the metric can be written as
\begin{equation}\label{LRSmetric}
 ds^2 = -dt^2 + A^2(t)\,(\omega^1)^2 + B^2(t)\,[(\omega^2)^2 +
(\omega^3)^2]\:,
\end{equation}
where the $\omega^i$ are invariant Bianchi type IX one-forms. We
consider models whose matter content is non-tilted dust with energy
density $\rho$, and a positive cosmological constant $\Lambda$. We
set $c=1=8\pi G$. The Hamiltonian for this class of models reads
\begin{equation}\label{H1}
H = \frac{p_A\,p_B}{4B} - \frac{A\,p_A^2}{8\,B^2} + 2A -
\frac{A^3}{2B^2} -2\Lambda\,A\,B^2 - E_0\:,
\end{equation}
where $p_A$, $p_B$ are the momenta conjugate to $A$, $B$. The
constant $E_0$ corresponds to a matter term, $E_0 = 2 \rho A B^2 =
\mathrm{const}$, which follows from the contracted Bianchi
identities. Einstein's field equations are given by
\begin{subequations}\label{eq:hamevol1}
\begin{align}
\label{AB} H &= 0\:,\qquad \dot A = \frac{p_B}{4B} -
\frac{A\,p_A}{4B^2}\:,\qquad
\dot B = \frac{p_A}{4B}\:,\\
\label{pA} \dot p_A
& = \frac{p_A^2}{8B^2} - 2 + \frac{3A^2}{2B^2} + 2\Lambda B^2\:,\\
\label{pB} \dot p_B & = \frac{p_A\,p_B}{4B^2} - \frac{A\,p_A^2}{4B^3} -
\frac{A^3}{B^3} + 4\Lambda AB\:.
\end{align}
\end{subequations}

The system~\eqref{eq:hamevol1} admits Einstein's static universe as
a solution, represented by the fixed point $\text{E}$: $p_A = p_B
=0$, $A= B = 1/\sqrt{4 \Lambda}$, $E_0 = 1/\sqrt{4 \Lambda}$.
In~\cite{heietal05}, the field equations were reformulated in terms
of scale-invariant variables in order to obtain a regular system.
This allowed the use of rigorous dynamical systems methods and thus
resulted in a comprehensive description of the global dynamics; in
particular, the dynamics of models initially close to $\text{E}$ was
found to be predictable and non-chaotic. Since the results are
completely invariant and do not rely on the particular choice of
variables, the findings of~\cite{heietal05} refute earlier claims
about ``homoclinic chaos'' \cite{olietal97,baretal01,olietal02,soastu05}.

The claims about homoclinic chaos rely on the
assertion that there exist infinitely many solutions
of~\eqref{eq:hamevol1} that evolve from an initial state close to
$\text{E}$, leave any neighborhood of $\text{E}$, and return to a
final state close to $\text{E}$ again. It is argued that this
behavior is the basis of chaos in the dynamics because of the
chaotic distribution of ``homoclinic'' initial data points among
initial data close to $\text{E}$. In the
following we demonstrate that the asserted ``homoclinic solutions''
can only exist under unphysical assumptions.
For pedagogical reasons, in our considerations we focus
on the most recent publication of~\cite{olietal97,baretal01,olietal02,soastu05},
i.e.,~\cite{soastu05};
in particular we adopt the notation and terminology.%
\footnote{We also adopt the term
``homoclinic solutions'' here, although this is a deviation from
standard dynamical systems nomenclature.}

The analysis of~\cite{soastu05} is crucially based on the
observation that the system~\eqref{eq:hamevol1} is regular for all
$A$ including $A=0$. This prompts the authors to not restrict the
domain of the variable $A$ to $A>0$ (``truncation at $A=0$ is bound
to a loss of information''), but to consider the regions $A<0$ and
$A=0$ as a part of phase space (``for the Hamiltonian dynamics [the
region $A<0$] constitutes an essential part of the phase space'').

As a first comment, let us note that the regularity of the
equations~\eqref{eq:hamevol1} at $A=0$ is inseparably connected with
the assumption of LRS symmetry. Generalizing~\eqref{LRSmetric} to
the full diagonal type IX case with spatial metric
$A^2 (\omega^1)^2 + B^2 (\omega^2)^2 + C^2 (\omega^3)^2$
yields
\[
-\frac{1}{8} \left[ \frac{A p_A^2}{B C}+ \frac{B p_B^2}{A C}+ \frac{C p_C^2}{A B} \right]
+\frac{1}{4} \left[ \frac{p_A p_B}{C}+ \frac{p_A p_C}{B}+\frac{p_B p_C}{A}\right]
\]
for the kinetic term of the Hamiltonian. (For $B=C$, $p_B=p_C$, the
kinetic term of~\eqref{H1} is recovered by noting that the new
momentum $p_B$ differs from the original one by a factor of 2.) We
observe that the non-LRS Hamiltonian and the resulting equations are
singular at $A=0$ (and $B=0$, $C=0$). Any deviation from LRS thus
invalidates the statement that ``the regions of phase space $A>0$
and $A<0$ join smoothly in $A=0$.'' This will be further discussed
below.

In general relativity, for LRS line elements of the
form~\eqref{LRSmetric}, one usually makes the restriction $A>0$ (and
$B>0$); in particular, Hamiltonian approaches typically use
exponential representations of the scale factors, e.g., $A=
e^{\alpha}$, see,
e.g.,~\cite{grav73,dametal03,waiell97,ren95,col03}. It is tacitly,
and correctly, assumed that the restriction to positive scale
factors is no loss of generality. This is due to the existence of
discrete symmetries: the system~\eqref{eq:hamevol1} is invariant
under $(A,p_A,t)\mapsto -(A,p_A,t)$ and $(B,p_B)\mapsto -(B,p_B)$
(where the latter reflects the invariance of the non-LRS system
under the simultaneous transformations  $(B,p_B, t) \mapsto
-(B,p_B,t)$ and $(C,p_C,t)\mapsto -(C,p_C,t)\,$). As a consequence,
solutions with $A<0$ arise from solutions with $A>0$ via this
symmetry map; in this context it is important to note that $E_0
\mapsto -E_0$ and that $\rho$ is invariant under the
transformations.

The phase space considered in~\cite{soastu05} is the union of the
regions $A>0$ and $A<0$, ``smoothly joined on $A=0$''; in
particular, $A=0$ is not regarded as a singularity from a
Hamiltonian point of view. The authors give numerical evidence for
the existence of an infinite set of solutions with initial data
close to the fixed point $\text{E}$ (i.e., with $A>0$) that pass
through $A=0$, enter the region $A<0$, re-enter $A>0$ through $A=0$,
and eventually enter a neighborhood of $\text{E}$ again.
It is the existence of these ``homoclinic solutions'' that is argued
to cause ``homoclinic chaos'' in the model. For the homoclinic
solutions $E_0 =\mathrm{const} >0$ holds; in fact, $E_0$ is close to
$1/\sqrt{4 \Lambda}$, which is the value characterizing Einstein's
universe $\text{E}$.

The positivity of $E_0$ has far reaching
consequences: from $E_0 = 2 A B^2 \rho$ we obtain
\begin{equation}\label{rho}
\rho = \frac{E_0}{2\, A\,B^2}\:,
\end{equation}
which implies that $\rho < 0$, if $A<0$ (since $E_0> 0$). Therefore
models described by solutions with $A<0$ violate the
(weak/null/strong/dominant) energy condition; indeed, since the
matter content is dust, $\rho<0$ implies that the dust particles
have negative inertial and gravitational mass. The homoclinic
solutions described in~\cite{soastu05} evolve between the regions
$A>0$ and $A<0$, hence $\rho>0$ changes to $\rho<0$ (and back)
during the evolution. However, the change of the density of the dust
particles from positive to negative is not continuous: when $A
\searrow 0$ we find $\rho \rightarrow \infty$, when $A \nearrow 0$
we have $\rho \rightarrow -\infty$. (This is not immediate
from~\eqref{rho}, but follows from the fact that the spatial volume density
$A B^2$ converges to zero as the singularity is approached.) Hence
extending a solution from $A>0$ via $A=0$ to $A<0$ (and back)
corresponds to joining $\rho = +\infty$ and $\rho = -\infty$ (and
again $-\infty$ with $+\infty$).

In~\cite{soastu05} the authors emphasize that it is essential to
incorporate the regions $A<0$ and $A=0$ into the analysis, since the
``dynamical phenomena that occur in $A<0$ are present in $A>0$
transported there via the Hamiltonian flow''. Naturally, there exist
numerous examples in physics where the consideration of unphysical
states is used as an abstract mathematical tool that makes possible
or facilitates the analysis of physical states. Here the case is
totally different. The system~\eqref{eq:hamevol1} is an autonomous
system, which makes the flow a local phenomenon. In particular the
phase space picture in the region $A>0$ depends exclusively on
$A>0$; it is a basic principle that (changes of) the flow in $A<0$
cannot affect the phase space picture in $A>0$.

Though irrelevant, the assertion that the system~\eqref{eq:hamevol1}
in the region $A<0$ is associated with complex dynamics (see, e.g.,
Figs.~17 and~18 in~\cite{soastu05}) is likely to be correct, but
considering that we are dealing with dust particles with negative
masses this should perhaps not be too surprising. Note that the
occurrence of negative masses is not primarily due to $A<0$, but due
to the fact that $A<0$ while $E_0$ is kept positive. Recall in this
context that application of the symmetry map $(A, E_0) \mapsto
-(A,E_0)$, cf.~the discussion above, would leave the dynamics (and
the positivity of $\rho$) unchanged.

We conclude that there exist two attitudes one can take to
homoclinic chaos in LRS type IX models with dust and a positive
cosmological constant.
\begin{itemize}
\item[(i)]  A conservative attitude, which is
the prevailing opinion of the classical general relativity
community; this is also the view we hold. Restricting the phase
space to positive scale factors is no loss of generality; $A=0$
represents a singularity of the model, since $\rho \rightarrow
\infty$ when $A\rightarrow 0$. Extending a cosmological model as a
solution with $A<0$ (and $E_0>0$) amounts to joining $\rho=+\infty$
with $\rho=-\infty$, and it must be assumed that the dust particles
replace their originally positive mass by negative mass. Any such
extension is unreasonable from the point of view of physics, and
irrelevant for the description of physical states from the point of
view of mathematics.
\item[(ii)] An 
unconventional attitude.
The phase space is given by the union of $A>0$, $A=0$, and $A<0$;
$A=0$ is not a singularity in the sense that the
evolution of cosmological models does not stop. There exists a
mechanism that transports the model (which is characterized by $E_0 >0$
and thus has positive mass originally)
through $A=0$ ($\rho=\pm \infty$) to a model with negative mass, and
back. The constructed spacetimes are of the form $\mathbb{R} \times
M$, where the manifold $M$ is Riemannian except for at a discrete
set of times, where the metric on $M$ is degenerate. At times of
degeneracy, the density of the dust particles changes sign from
$\rho>0$ via $\rho = +\infty$ and $\rho =-\infty$ to $\rho <0$ (or
reversed).
\end{itemize}

If one takes view (i), it can be
proved, see~\cite{heietal05}, that no homoclinic chaos exists and
that the dynamics is predictable. The results in~\cite{heietal05}
are completely invariant and do not rely on the particular
scale-invariant formulation; however, the results do rely on the
assumption that the dust particles retain their positive mass.%
\footnote{The essence of view (i) pervades all theories that model
phenomena with differential equations: although the equations might be
well-defined (and regular) for a range of a variable that is greater than its
physical range, the physicality condition prohibits the extension of
solutions. Examples abound not only in physics; for instance, in
population models in mathematical biology~\cite{britton},
solutions will never be extended to negative numbers of individuals of a species.}

View (ii) is associated with severe problems. First, we note that
although formally, i.e., from a purely mathematical point of view,
the extension of the scale factor $A$ through $A=0$ is feasible, this
extension relies crucially on the assumption of LRS symmetry. When
LRS symmetry is broken (as it certainly is for generic models) it is
far from clear whether an extension of generic solutions through the
singularity exists. Let us elaborate on this issue.

In the LRS case, the regularity of~\eqref{H1}
and~\eqref{eq:hamevol1} in $A$ allowed~\cite{soastu05} to also
regularize the problem in $B$ by a change of the time variable that
only involved $B$. In this formulation, $B=0$ was an invariant
subset, so that orbits could not pass through $B=0$; in contrast,
the set $A=0$ was not invariant, which was exploited to extend
orbits beyond the singularity $A=0$. In the full diagonal type IX
case the equations are singular at $A=0$, $B=0$, and $C=0$. It is
possible to regularize the equations by a change of the time
variable (in a spirit similar to the $B$ regularization
in~\cite{soastu05}). However, this re-parametrization must
necessarily involve all three scale factor due to permutation
symmetry, e.g., one could choose a lapse proportional to $A B C$. As
a consequence, $A=0$, $B=0$, and $C=0$ become invariant subsets,
which prevents solutions from changing the sign of the scale factors
(and thus from changing the sign of $\rho$). Hence, in the non-LRS
case, an extension of solutions through the singularity lacks a
mathematical foundation.

That LRS models are of a special nature is well known. In
particular, singularities that occur in the LRS case are very
different from those occurring in the diagonal case. In the LRS case
solutions generically approach the singularity along a so-called
Taub asymptote. This suggests that the singularity may be a weak
null singularity, see~\cite{limetal05} and references therein, and
that is possible to extend the spacetime through the singularity in
a $C^0$ manner to the Minkowski spacetime. It is thus not
inconceivable that the metric can be extended, but it is unclear
whether such a construction bears any relation to the extension of
the scale factor $A$ 
critically discussed above.
The problem of
whether the extension of $A$ leads to an extension of the metric is
not discussed in~\cite{soastu05}; clearly, the extension of $A$ is
completely irrelevant unless it leads to an extension of the metric.

In the diagonal (non-LRS) case, on the other hand,
the situation is completely different: solutions
generically approach the singularity in a Mixmaster fashion and the
singularity is a spacelike scalar curvature
singularity~\cite{rin01}; no extension of the metric is possible.
Since the Mixmaster behavior is typical for generic cosmological
singularities, see e.g.~\cite{uggetal03} and references therein,
these considerations seem to rule out the type of manipulations and
arguments presented in~\cite{soastu05} for generic models. It
appears that the special LRS case is quite misleading when it comes
to spacetimes and phase space extensions.

Even in the LRS case, not all solutions possess extensions of the
kind discussed above; 
a Taub asymptote is a
prerequisite. This is because for such solutions $B\not \rightarrow
0$ when the singularity is approached; solutions with $B\rightarrow
0$, e.g. solutions with an isotropic singularity, are inextendible.
This creates a link to the related question of why it is possible
for unphysical solutions ($\rho<0$) to expand from a singularity,
and then recollapse to a singularity at $A=0$ (and subsequently
re-enter the physical part of the phase space). At first this is
quite surprising since dust particles with negative mass generate
anti-gravity, and one would thus expect such solutions to expand to
a state infinite dilution, $\rho \nearrow 0$, which is confirmed by
numerics for typical solutions. However, the solutions found
numerically in~\cite{soastu05} have Taub asymptotes and are thus
vacuum dominated with extreme shear; since shear represents gravity
generating gravity, these solutions can recollapse to the
singularity and enter the physical region again. Note, however, that
this scenario is only correct for solutions with Taub asymptotes
with initial data such that the shear dominates during a
sufficiently long time.

Finally, let us point out that, irrespective of the mathematical
considerations, at present there does not exist any physical theory
that could explain the traversing of the singularity in the way
implicitly suggested by adherents of view (ii). 
Note that certain speculative
mechanisms like ``quantum tunneling of the wave function of the
universe'' (see~\cite{jorstu03}) are insufficient, since they lack
an explanation for the metamorphosis of the dust particles from
positive to negative mass. In addition, in view of the discussion
above, such a mechanism is a serious candidate only if it can
explain the behavior of generic models and not merely the
exceptional LRS type IX models.

It seems to us that the burden of making a convincing case for
homoclinic chaos in Bianchi type IX models lies with adherents to
view (ii); considering the above, this is a formidable challenge indeed.

\vspace{2ex}

\begin{center}
\textit{Note added.}
\end{center}

After submission of this letter an erratum to~\cite{soastu05} was
published, see~\cite{soastu05E}. In this small note the authors
state that the physical interpretation on the existence of
homoclinic chaos has to be corrected, but the authors do not give
any explanation for this statement (see, however, the first part of
the present letter); furthermore, it is not discussed that the
corrected interpretation also affects the status of homoclinic chaos
in non-LRS models (see, however, the second part of the present
letter).

\bibliography{IXlett}

\end{document}